\documentclass[aps,prl,twocolumn]{revtex4}
\usepackage[usenames]{color}
\newcommand{\comment}[1]{}
\usepackage{graphicx}
\usepackage{epsfig}
\usepackage{amsmath}
\usepackage{mathrsfs}
\newcommand{\bra}[1]{\langle {#1} |}
\newcommand{\ket}[1]{| {#1} \rangle}
\newcommand{\expect}[1]{\langle {#1} \rangle}
\newcommand{\ketn}[1]{ {#1} \rangle}
\newcommand{\bo}{{\bf \Omega}}
\usepackage{color}

\begin{document}
\title{Antiferromagnetic Spinor Condensates are Quantum Rotors}
\date{\today}
\author{Ryan Barnett, Jay D. Sau, and S. Das Sarma}
\affiliation{Joint Quantum
Institute and Condensed Matter Theory Center, 
Department of Physics, University of Maryland, College
Park, Maryland 20742-4111, USA}

\begin{abstract}
We establish a theoretical correspondence between spin-one
antiferromagnetic spinor condensates in an external magnetic field and
quantum rotor models in an external potential.  We show that the rotor
model provides a conceptually clear picture of the possible phases and
dynamical regimes of the antiferromagnetic condensate.  We also show
that this mapping simplifies calculations of the condensate's spectrum
and wavefunctions. We use the rotor mapping to describe the different
dynamical regimes recently observed in $^{23}$Na condensates
\cite{liu09a,liu09b}.  We also suggest a way to experimentally observe quantum
mechanical effects (collapse and revival) in spinor condensates.
\end{abstract}
\maketitle

 \vspace{.5in}

Bose-Einstein condensates occurring in ultracold atoms having internal
spin degrees of freedom, the so-called spinor condensates, 
offer an exciting addition to the family of quantum many body spin
systems realizable in the laboratory \cite{stenger98}.  Of particular
interest are the long coherence times and small dissipation rates
which allow access to dynamical regimes not available in the solid
state.  Recently there has been considerable experimental progress in
elucidating the dynamics of spinor condensates. Such endeavors include
dynamics experiments on $^{87}$Rb atoms for the hyperfine spin-one
\cite{chang05, sadler06} and spin-two \cite{klempt09a} manifolds as
well as, 
most recently, experiments on
$^{23}$Na condensates \cite{liu09a,liu09b}.  $^{23}$Na spin-one
condensates are qualitatively different than their $^{87}$Rb
counterpart due to antiferromagnetic interactions.  This leads to
ground states having zero spin moment as well as disparate dynamical
regimes.  

The NIST experiments \cite{liu09a,liu09b} were performed in a trapping
potential sufficiently tight such that, within a good approximation,
the bosonic atoms all occupy the same spatial mode.  This allows the
spin dynamics of the system, which is often obscured by spatial
variations, to be directly probed.  The condensate was prepared in an
initial unstable ferromagnetic state and then allowed to evolve freely
in time.  For small magnetic fields, the system oscillates about the
ferromagnetic state, never reaching zero spin moment $\expect{\bf F}=0$
at any time.  On the other hand, when the magnetic field exceeds a
critical value the system evolves through $\expect{\bf F}=0$ reaching
a state pointing in the opposite direction and back periodically, the
so-called ``running phase'' trajectories.   It was shown that these
different regimes could be interpreted as being on different sides 
of a separatrix in the phase
space of the mean-field energy of the system \cite{liu09a,liu09b}.

In the single-mode approximation, the full quantum Hamiltonian of the
system is
\begin{equation}
\label{Eq:H} H = \frac{g}{2N} F^2 - q a_0^\dagger a_0.
\end{equation} 
Here, ${\bf F}=a^\dagger_{\alpha} {\bf F}_{\alpha\beta}a_{\beta}$ is
the total spin operator where ${\bf F}_{\alpha \beta}$ are the
spin-one matrices and $a_{1}, a_{0}, a_{-1}$ are bosonic annihilation
operators for each spin state, $N$ is the total particle number, $g$
is the spin-dependent interaction, and $q$ is the quadratic Zeeman
shift due to an external magnetic field 
\cite{fn}.  When $q=0$ the exact ground state of the above
Hamiltonian is a condensate of singlet pairs of bosons given by
\cite{law98}
\begin{equation} 
\label{Eq:singlet}
\ket{S}= \left(a_0^\dagger a_0^\dagger - 2
a_1^\dagger a_{-1}^\dagger\right)^{N/2} \ket{0}.
\end{equation} 
This ground state is unique and breaks no symmetries.
However, for large particle numbers, this state becomes extremely
delicate, being unstable to small external magnetic fields.  
Thus the observed phases for most experimental antiferromagnetic systems
are more appropriately described by symmetry-broken nematic states
which are well-described by mean-field theory
\cite{ho98,ohmi98}.  This is reminiscent of  Anderson's ``tower
of states''  argument for N\'{e}el ordering in solid state 
quantum antiferromagnets, despite the fact that the true ground state
for finite-size bipartite lattices can be shown to be a spin singlet \cite{auerbach94}.

In this Letter, we develop a conceptually new approach to describe the
quantum dynamics of
antiferromagnetic spinor condensates.  In
particular, we map the Hamiltonian in Eq.~(\ref{Eq:H}) onto a quantum
rotor Hamiltonian
\begin{equation}
\label{Eq:Hr} 
{\mathscr H}= \frac{1}{2I} L^2 + V(\theta)
\end{equation} 
where ${\bf L}$ is the angular momentum of the rotor,  
$I=N\hbar^2/g$ is the moment of inertia, and 
$V(\theta) = q\left(N+\frac{3}{2}\right)
\sin^2(\theta) + \frac{q^2N}{8 g}\sin^2(2\theta)$ 
is the external potential.  The mapping is exact
in the sense that the complete spectrum of Eq.~(\ref{Eq:H}) for $N$
bosons precisely agrees with the lowest set of eigenvalues of
Eq.~(\ref{Eq:Hr}) (which has an unbounded spectrum from above).  
A similar procedure has been used to derive an exact phase model
describing bosons in a double-well potential \cite{anglin01}.
One can see that the singlet state of paired bosons, Eq.~(\ref{Eq:singlet}),
corresponds to a state where the rotor is delocalized over the entire sphere
while the symmetry-broken nematic state corresponds to the rotor being
in a position eigenstate.  We will show
how Eq.~(\ref{Eq:Hr}) can be used to obtain simple expressions for the
spectrum and wavefunction of the spinor condensate.
We then show how the semiclassical limit of the rotor system provides
a natural interpretation of the dynamical regimes of anti-ferromagnetic
spinor condensates observed experimentally \cite{liu09a, liu09b}.  
Finally, we make a prediction to observe quantum
mechanical effects (i.e. non-mean field effects) in spinor condensates
which have so far eluded experimental detection.  Specifically, we
show that the abrupt removal of a magnetic field used to prepare the
system in a nematic state will lead to collapse and revival dynamics, which
cannot be explained with mean field theory alone.

We now proceed with the main technical advance of this work: an exact
mapping of Eq.~(\ref{Eq:H}) onto an effective rotor Hamiltonian, thus
establishing that antiferromagnetic spinor condensates are effective
realizations 
of the quantum rotor model.  It is most useful to express the bosonic
creation and annihilation operators as quantities that transform as
cartesian vectors under rotations.  To that end we define the
operators
$
b_x = -(a_1 - a_{-1})/\sqrt{2},
$
$
b_y  = (a_1 + a_{-1})/i\sqrt{2},
$
and
$
b_z = a_0
$
which satisfy bosonic commutation relations.  It is then
straightforward to express the Hamiltonian Eq.~(\ref{Eq:H}) in terms
of these operators.  Specifically, the spin operator is ${\bf F}= -i
{\bf b}^\dagger \times {\bf b}$ while the quadratic Zeeman shift is
$b_z^\dagger b_z$.  With these operators we construct the complete set
of states
\begin{equation}
\label{Eq:omega} 
\ket{\bo_N} \equiv \frac{1}{\sqrt{N!}} \left( \bo
\cdot {\bf b} ^\dagger\right)^N \ket{0}
\end{equation} where $\bo=(\sin(\theta)\cos(\phi), \sin(\theta)
\sin(\phi), \cos(\theta))$ is a real unit vector given by the pair of spherical
coordinates $(\theta,\phi)$ and $N$ is the number of bosons in the
system. For simplicity we take $N$ to be even and will comment on the
odd $N$ case shortly.  This wavefunction is
the (symmetry broken) nematic state pointing along $\bo$.   
These states have the inner
product
\begin{equation}
\label{Eq:inner}
\bra{\bo_N} \ketn{\bo'_N} = \left( \bo \cdot \bo' \right)^N.
\end{equation}
Thus, as the number of bosons in the system becomes large, 
states pointing in different directions become orthogonal.

Interestingly, the spin-singlet state Eq.~(\ref{Eq:singlet}) can be constructed by taking
equal-weight superpositions of the nematic state over all directions:
\begin{equation}
\int d\Omega  \ket{\bo_N} \propto
\left(a_0^\dagger a_0^\dagger - 2 a_1^\dagger
  a_{-1}^\dagger\right)^{N/2} \ket{0} = 
\left( {\bf b}^\dagger \cdot {\bf b}^\dagger \right)^{N/2} \ket{0}
\end{equation}
as discussed in Refs.~\cite{ashhab02,mueller06}.  This motivates
one to use the spherical harmonics to construct the orthonormal set of
states
for even $\ell$:
\begin{equation}
\ket{\ell m} = \frac{1}{\sqrt{f_\ell}}\int d\Omega Y_{\ell m}(\bo) \ket{\bo_N}
\end{equation}
where 
$
f_\ell = 4 \pi N! 2^\ell \left(\frac{N+\ell}{2}\right)!/
\left(\frac{N-\ell}{2}\right)!(N+\ell+1)!
$ is the normalization constant.  
Such states are defined for
$|\ell|\le N$, and unless otherwise stated sums for over such states are
understood to satisfy this restriction.
These states
$\ket{\ell m}$ can be seen to be eigenstates of the $F^2$ operator 
with eigenvalue $\ell(\ell+1)$.  We finally note that these have 
the following inner product with the nematic states
\begin{equation}
\label{Eq:prod}
\bra{\bo _N} \ketn{\ell m} = \sqrt{f_\ell} Y_{\ell m} (\bo).
\end{equation}

With the construction of these two sets of basis states $\ket{\bo_N}$
and $\ket{\ell m}$ in the bosonic Hilbert space we now proceed to map
the problem onto the rotor Hilbert space.  This Hilbert space is
spanned by the position eigenstates $\ket{\bo}$ on the unit sphere
which are complete and satisfy the orthonormality condition
$\bra{\bo}\ketn{\bo'}=\delta(\bo-\bo')$. 
These states involve angular
momentum components for all $\ell$ and therefore do
not suffer the complications that arise from Eq. (4) for
the $\ket{\bo_N}$ states which are only orthogonal in the large $N$
limit .
To start we note that a general state in the bosonic Hilbert
space can be written as a superposition of the spin nematic states
with weight $\psi(\bo) = \bra{\bo}\ketn{\psi}$:
\begin{equation}
\label{Eq:psispectral}
\ket{\Psi} = \int d\Omega \ket{\bo_N} \psi(\bo).
\end{equation}
We now act with $H$ on this state.
If one can find an operator ${\cal H}$ in the rotor Hilbert space 
such that
\begin{equation}
\label{Eq:spectral}
H \ket{\Psi} = \int d\Omega \ket{\bo_N} \bra{\Omega}  {\cal H} \ket{\psi}
\end{equation}
then a sufficient condition for the time-dependent Schrodinger
equation (TDSE) in the bosonic Hilbert space to be satisfied is the
rotor TDSE: ${\cal H} \ket{\psi} = i \hbar \partial_t \ket{\psi}$. 
The necessary condition for the rotor model to be a precise
description for spinor condensates may be less restrictive.

Our efforts will now be devoted to 
showing that ${\cal H}$ exists and then finding ${\cal H}$.  We 
consider the two terms of the bosonic Hamiltonian  
Eq.~(\ref{Eq:H}) separately.  The first term, which contains
$F^2$, is diagonal in the
$\ket{\ell m}$ representation which simplifies the mapping.
It is intuitive that $F_{\alpha}$ will map to the angular momentum
operator in the rotor Hilbert space defined as 
$L_\alpha = -i \hbar \varepsilon_{\alpha \beta \gamma} \Omega_\beta \nabla_\gamma$.  
This can be derived by inserting the completeness relations
$1=\sum_{\ell m} \ket{\ell m} \bra{\ell m}$ and $1=\int d\Omega
\ket{\bo}\bra{\bo}$ (which act in different Hilbert spaces).  Using
Eq.~(\ref{Eq:prod}) we obtain
\begin{align}
F^2 \ket{\Psi} &= 
\sum_{\ell m} \int d\Omega F^2 \ket{\ell m } 
\sqrt{f_\ell} \bra{Y_{\ell m}}\ketn{\bo} \bra{\bo}  \ketn{\psi}
\notag
\\
&= \frac{1}{\hbar^2}\sum_{\ell m} \ket{\ell m } 
\sqrt{f_\ell} \bra{Y_{\ell m}} L^2 \ket{\psi}
\notag
\\
&=
\frac{1}{\hbar^2}
\int d\Omega \ket{\bo_N} \bra{\bo} L^2 \ket{\psi}
\end{align}
where we have used the notation 
$\bra{\bo}\ketn{Y_{\ell m}} \equiv Y_{\ell m}(\bo)$.  Thus we see that
\begin{equation}
\label{Eq:Fmap}
F^2 \rightarrow \frac{1}{\hbar^2} L^2
\end{equation}
in the rotor representation.  Such a rotor description of $F^2$ was
previously noted in \cite{zhou01,demler02,imambekov03}.

We now move on to mapping the quadratic Zeeman term in $H$ to a rotor
description.  This mapping is more complicated since the quadratic
Zeeman shift is not diagonal in either the $\ket{\bo_N}$ or the
$\ket{\ell m}$ representation.  Our approach will be to express
$b_z^\dagger b_z \ket{\bo_N}$ in terms of $\ket{\bo_N}$ and its
derivatives.  Then integration by parts can be used to arrive at
Eq.~(\ref{Eq:spectral}).  In the analysis we consider general
quadratic terms of the form $b_{\alpha}^\dagger b_\beta$.  We state
without derivation the following identity
\begin{equation}
\label{Eq:identity}
b_{\alpha}^\dagger b_{\beta} \ket{\bo_N} = \Omega_\beta \left( 
\nabla_\alpha+ N \Omega_\alpha
\right) \ket{\bo_N}
\end{equation}
where ${\bf \nabla}=\hat{\theta}\partial_{\theta} +
\frac{1}{\sin(\theta)} \hat{\phi} \partial_{\phi}$ is the gradient operator on the
unit sphere.  
This identity  
follows from the geometrically intuitive relation
$
\nabla_\alpha \Omega_\beta= \delta_{\alpha \beta} -
\Omega_\alpha \Omega_\beta.
$
We finally note that the integration by parts rule for $\nabla_\alpha$
is \begin{equation}
\label{Eq:parts}
\int d\Omega \; f (\bo) \nabla_\alpha g (\bo) 
= \int d\Omega \; g (\bo) \left[ 2\Omega_\alpha - 
\nabla_\alpha \right] f (\bo). 
\end{equation}

Using Eqns.~(\ref{Eq:psispectral}), (\ref{Eq:identity}), and (\ref{Eq:parts})
we obtain
\begin{align}
\label{Eq:quadratic}
b_\alpha^\dagger b_\beta \ket{\Psi} &= 
\int d\Omega \psi(\bo) \Omega_\beta \left( 
\nabla_\alpha+ N \Omega_\alpha
\right) \ket{\bo_N}
\\
\notag
&=
\int d\Omega \ket{\bo_N}
\left( 
(N+3) \Omega_\alpha \Omega_\beta - \Omega_\beta
\nabla_\alpha - \delta_{\alpha\beta} 
\right) \psi(\bo).
\end{align}
From this we can read off the equivalent operator acting in the rotor space
which corresponds to $b_\alpha^\dagger b_\beta$:
\begin{equation}
\label{Eq:QZmap}
b_\alpha^\dagger b_\beta \rightarrow  
(N+3) \Omega_\alpha \Omega_\beta - \Omega_\beta
\nabla_\alpha - \delta_{\alpha\beta} .
\end{equation}
Using the mappings in (\ref{Eq:Fmap}) and (\ref{Eq:QZmap})
restricted to the case $\alpha = \beta = z$,
we finally arrive at
the operator ${\cal H}$:
\begin{equation}
{\cal H}  = \frac{g}{2N\hbar^2} L^2 -  q(N+3)\Omega_z^2 + q \Omega_z
\nabla_z
\end{equation}
where $\nabla_z= -\sin(\theta) \partial_\theta$ and we have dropped a constant term.
While ${\cal H}$ has a real spectrum, it is not  Hermitian.  It is
therefore advantageous to apply a similarity transformation
to render it Hermitian.  Defining
\begin{equation} 
{\mathscr H} = e^{F} {\cal H}
e^{-F}\end{equation}
with $F=-\frac{qN}{4g}\cos(2\theta)$ we arrive at
Eq.~(\ref{Eq:Hr}) and the mapping is complete.  We note that with
this transformation, the wavefunctions $\psi(\bo)$ governed
by ${\mathscr H}$, when  entering
Eq.~(\ref{Eq:psispectral})  must be accompanied by a factor
of $e^{-F}$.

This equation is the model for a quantum rotor under an external
potential.  Since we are taking the case of even $N$ the wavefunctions
must satisfy the constraint $\psi(\bo) = \psi(-\bo)$.  This condition
can be interpreted as constraining the ends of the rotor to be bosonic
particles, requiring the rotor wavefunction to be symmetrical 
under their interchange.
This constraint can be enforced with the projection operator
$
{\cal P} = \sum_{{\rm even} \; \ell} \sum_{m=-\ell} ^ {\ell}
\ket{Y_{\ell m}} \bra{Y_{\ell m}}.
$
Since this operator commutes with the Hamiltonian Eq.~(\ref{Eq:Hr})
the constraint imposes no real technical difficulty.  The case of odd
$N$ is similar  and is therefore not shown here. 
For this the wavefunction must be antisymmetric and
the corresponding projection operator running over odd $\ell$ will
also commute with the Hamiltonian.

We now consider the limiting cases of the rotor Hamiltonian.
The simplest situation is when no external magnetic field is present and $q=0$.
For this the ground state is uniformly delocalized over the entire
sphere corresponding to the $\ell=m=0$ spherical harmonic. 
We now consider the case of small magnetic field such that
$g \gg q >0$.   For this case the first term in the rotor
potential $V(\theta)$ dominates  and serves to localize the rotor
about the poles.  In this limit, we can expand the potential to quadratic
order about the $\theta=0$ minimum and the Hamiltonian becomes that
of a two-dimensional harmonic oscillator \cite{cui08}.  The spectrum for the lowest energies are
then 
\begin{equation}
\label{Eq:HOspec}
\varepsilon_n=\sqrt{2gq}(n+1)
\end{equation}
(for even $n$ with multiplicity $2n+1$) and the ground state
wavefunction is
\begin{equation}
\label{Eq:howf}
\psi_0(\theta) = \sqrt{\frac{1}{\pi\bar{\theta}^2}}e^{-\theta^2/2\bar{\theta}^2}
\end{equation}
where the oscillator length is $\bar{\theta}=\sqrt{
\frac{g}{2 qN^2}}$.  That the energy states are
evenly spaced and have the spectrum given by Eq.~(\ref{Eq:HOspec}) in
this regime is not immediately clear from a direct analysis of the
original bosonic Hamiltonian Eq.~(\ref{Eq:H}).  In order for this
harmonic oscillator description to be valid we must have the condition
$\bar{\theta} \ll 1$.  Away from this limit the rotor will delocalize
and approach the singlet state.  For a large particle number $N$ we
therefore see that any small external magnetic field will tend to
drive the system to the symmetry broken nematic state as described by
the mean field theory \cite{ho98,ohmi98}.  For higher magnetic field
we see that when $q>2g$ a local minimum appears along the equator
$\theta=\frac{\pi}{2}$ though the global minimum will remain at
$\theta=0$.  This leads to stationary states localized about the
equator.  Such states are analogous to the
``$\pi$-states'' occurring for a scalar condensate in a double-well
potential \cite{raghavan99}.  However, as in the double-well
case,  transforming this wavefunction back to the bosonic Hilbert space
can significantly alter its structure \cite{anglin01}.

Having described quantum mechanical 
states of Eq.~(\ref{Eq:Hr}) in various limiting cases  we now proceed
to a semi-classical analysis of its dynamics which is relevant to
the recent experimental results \cite{liu09a,liu09b}.  The 
Lagrangian describing the motion in the semiclassical limit is
\begin{equation}
{\cal L} = \frac{1}{2} I \left(  \dot{\theta}^2  +
 \sin^2(\theta) \dot{\phi}^2 \right) - V(\theta).
\end{equation}
The equation of motion for this is
\begin{equation}
I\ddot{\theta} =
I \frac{\cos(\theta)}{\sin^3(\theta)}p_\phi^2-\frac{\partial V}{\partial \theta}
\end{equation}
where 
$p_{\phi} =  \sin^2(\theta) \dot{\phi}$ is a constant of motion.  As
before we start by considering the limiting case $g\gg q >0$.  For
this case we can drop the second term in $V$.  The first type of
motion we consider is when the rotor remains close to the minimum
at the poles at all times.  The potential can then be expanded to
quadratic order in $\theta$ and analytic solutions can be found.
One solution is where the
rotor oscillates through the poles: $\theta(t)=\theta_0
\cos(\omega t)$, $\dot{\phi}=0$.  
Another solution is where the rotor
precesses about the poles:  $\dot{\theta}=0$, $\phi(t) = \omega t$.
Both of these solutions have the eigenfrequency $\omega=\sqrt{2gq}/\hbar$ which
corresponds to the energy scale appearing in the spectrum from the
quantum mechanical analysis Eq.~(\ref{Eq:HOspec}).
The second type of motion we consider is where
the rotor has enough energy to overcome the potential
barrier near the equator and explore both hemispheres in its trajectory.
These are precisely the oscillating phase solutions experimentally
observed in \cite{liu09a,liu09b}.   Finally, a third type of motion
is possible when $q>2g$.  As described above, for this case there is
a local minimum at the equator.  Therefore for this situation there
will  be trajectories which remain localized about the equator.

\begin{figure}
\includegraphics[width=3.2in]{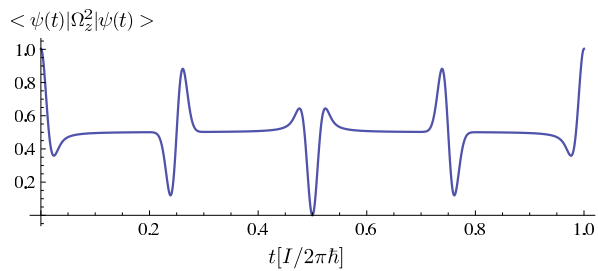}
\caption{Collapse and revival of $\bra{\psi(t)} \Omega_z^2
\ket{\psi(t)}$ starting from a state localized about
poles with width $\bar{\theta} = 0.1$.}
\label{Fig:revival}
\end{figure}

We now apply the rotor description to the quantum dynamics of
antiferromagnetic condensates in the single mode regime, which is 
known to manifest rich behavior \cite{romano04,diener06, zhai09}.  Here we consider
preparing the system in the
symmetry-broken nematic state given by Eq.~(\ref{Eq:howf}), and then
rapidly turning the magnetic field off and allowing the state to
evolve freely.  We note that according to the semiclassical theory
(or by using the Gross-Pitaevskii equation) the nematic wavefunction
will remain at the pole and not evolve temporally.  The quantum mechanical
dynamics, however, is markedly different.  By dynamically evolving the 
wavefunction, Eq.~(\ref{Eq:howf}), with the quantum rotor Hamiltonian
Eq.~(\ref{Eq:Hr}) with $q=0$, it can be seen that the state will
undergo periodic collapse and revival at the characteristic frequency
$\hbar /I$.  For instance, provided the initial state is sufficiently
localized ${\bar \theta}\ll 1$, one can show that
\begin{equation} 
\label{Eq:collapse}
\bra{\psi(t)} \Omega_z^2 \ket{\psi(t)} = 2
\bar{\theta}^2 \sum_{{\rm odd} \; \ell >0} (2\ell+1)
e^{-\ell(\ell+1)\bar{\theta}^2} \cos^2\left( \frac{(2\ell +1)\hbar
t}{2I}\right).
\end{equation}
The evolution of this function over a single period is plotted in
Fig.~\ref{Fig:revival}.  The localized nematic state rapidly
collapses to  states with substantial weight contributions from other
regions of the unit sphere, and then fully revives at the end of the period.
By applying the Poisson resummation formula to Eq.~(\ref{Eq:collapse}) 
it can be seen that the evolution is a train of localized pulses
separated by a fourth of the time period. 
This behavior can be  directly seen
experimentally by measuring the time dependence of $\expect{a_0^\dagger a_0}$
after the turning off the magnetic used to prepare the system
in the polar state. We note that since the magnetic field couples only to the spin degrees
of freedom, the above procedure will not excite spatial modes of the
condensate for sufficiently tight traps.  The quantum collapse and revival of
Fig.~\ref{Fig:revival} is a direct consequence of the rotor mapping of spinor condensates.

In conclusion, we have established a correspondence between
antiferromagnetic spinor condensates and quantum rotors.  We have
shown that this  mapping offers a considerable conceptual as well as
technical advance in understanding the properties of spinor
condensates.  We use the mapping to address recent experimental results
\cite{liu09a, liu09b} and to analytically predict a collapse and
revival process (which is a direct experimental signature of quantum
effects).  We point out that it should be
possible to provide similar quantum rotor mappings for condensates with larger
spin.

We would like to acknowledge insightful discussions with S. Maxwell and P. Lett.
This work was supported by JQI-NSF-PFC, DARPA QuEST, AFOSR, and ARO-DARPA-OLE.


\begin{thebibliography}{22}
\bibitem[{fn()}]{fn}
\bibinfo{note}{In terms of microscopic parameters, $g=\frac{4\pi
  \hbar^2}{3m}(a_2 - a_0)n_0$ where $m$ is the mass of the constituent atoms,
  $a_0$ and $a_2$ are scattering lengths, and $n_0$ is the equilibrium density
  of the condensate.}
\end{thebibliography}

\end{document}